\documentclass[preprint,proceedings]{rmaa}

\usepackage{paralist}

\usepackage{psfrag,color}

\newcommand{\BP}{Ballesteros-Paredes}

\newcommand{\Jc}{J_{\rm c}}

\newcommand{\Ms}{M_{\rm s}}
\newcommand{\muc}{\mu_{\rm c}}
\newcommand{\navg}{\bar n}
\newcommand{\npeak}{n_{\rm p}}
\newcommand{\nt}{n_{\rm t}}
\newcommand{\tfc}{\tau_{\rm fc}}
\newcommand{\tfg}{\tau_{\rm fg}}
\newcommand{\ts}{\tau_{\rm s}}
\newcommand{\VS}{V\'azquez-Semadeni}

\SetYear{2002}
\SetConfTitle{Gravitational Collapse: From Massive Stars to Planets}

\title{The Lifetimes of Molecular Cloud Cores: What is the Role of the
Magnetic Field?}

\author{
  E. V\'azquez-Semadeni,\altaffilmark{1}
  J. Kim,\altaffilmark{2}
  M. Shadmehri,\altaffilmark{3}
  and J. Ballesteros-Paredes\altaffilmark{1}}

\altaffiltext{1}{Centro de Radioastronom\'ia y Astrosf\'isica, UNAM,
Morelia, M\'exico.}
\altaffiltext{2}{Korea Astronomy Observatory, Korea.}
\altaffiltext{3}{Ferdowsi University, Mashhad, Iran.}

\shortauthor{V\'azquez-Semadeni et al.}
\shorttitle{Molecular Cloud Core Lifetimes}

\fulladdresses{
\item Enrique V\'azquez-Semadeni and Javier Ballesteros-Paredes:
Centro de Radioastronom\'ia y Astrosf\'isica, UNAM, Apdo. Postal
3--72,  58090 Morelia, Michoac\'an, M\'exico
(\email{e.vazquez,j.ballesteros@astrosmo.unam.mx}).
\item Jongsoo Kim: Korea Astronomy Observatory, 61-1, Hwaam-Dong,
Yusong-Ku, Taejon 305-348, Korea (\email{jskim@kao.re.kr}).
\item Mohsen Shadmehri: Department of Physics, School of Science, Ferdowsi
University, Mashhad, Iran (\email{mshadmehri@science1.um.ac.ir}).
}

\listofauthors{E.\ V\'azquez-Semadeni, J. Kim, M. Shadmehri \&
J. Ballesteros-Paredes}
\indexauthor{V\'azquez-Semadeni, E.}
\indexauthor{Kim, J.}
\indexauthor{Shadmehri, M.}
\indexauthor{Ballesteros-Paredes, J.}

\abstract{We discuss the lifetimes and evolution of dense cores formed
as turbulent density fluctuations in magnetized, isothermal molecular
clouds. We consider numerical
simulations in which we measure the cores' magnetic criticality and
Jeans stability in relation to the magnetic criticality of their
``parent clouds'' (the numerical boxes). In subcritical boxes, dense
cores do not form, and collapse does not occur. In supercritical
boxes, some cores collapse, being part of larger clumps that are
supercritical from the start, and whose central, densest regions
(the cores) are initially subcritical, but rapidly become
supercritical, presumably by accretion along field lines. Numerical
artifacts are ruled out. The time scales for cores to go from
subcritical to supercritical and then collapse are a few times the
core free-fall time, $\tfc$.
Our results suggest that cores are out-of-equilibrium, transient
structures, rather than quasi-magnetostatic configurations. }

\resumen{Discutimos los tiempos de vida y la evoluci\'on de los
n\'ucleos densos formados como fluctuaciones turbulentas de densidad
en nubes moleculares isot\'ermicas magnetizadas. Consideramos
simulaciones num\'ericas en las que medimos la criticalidad
magn\'etica y la estabilidad de Jeans de los n\'ucleos, en relaci\'on
a la criticalidad magn\'etica de sus nubes ``madre'' (los dominios de
integraci\'on num\'erica). En dominios subcr\'iticos, no se forman
n\'ucleos densos, y no ocurre colapso gravitacional. En dominios
supercr\'iticos, algunos n\'ucleos se colapsan, formando parte de
grumos m\'as grandes que son supercr\'iticos desde el inicio, y cuyas
regiones m\'as densas (los n\'ucleos) son inicialmente subcr\'iticas,
pero r\'apidamente se tornan supercr\'iticos, presumiblemente por
acreci\'on a lo largo de las l\'ineas de campo magn\'etico. Descartamos
la posibilidad de artefactos num\'ericos. Las escalas de tiempo en las
que los n\'ucleos transitan desde el estado subcr\'itico, pasando por
el supercr\'itico, hasta finalmente el colapso, son de unas cuantas
veces su tiempo de ca\'ida libre, $\tfc$. Nuestros resultados sugieren
que los n\'ucleos son estructuras transientes, fuera de
equilibrio, y no configuraciones cuasi-magneto-hidrost\'aticas.
}

\addkeyword{ISM: Molecular clouds}
\addkeyword{ISM: Turbulence}
\addkeyword{Stars: Formation}

\begin{document}
\maketitle

\section{Introduction} \label{sec:intro}

The prevailing view (which we hereafter refer to as the ``standard
(magnetic support) model'' of star formation; see, e.g., the
reviews by Shu, Adams \& Lizano 1987; McKee et al.\ 1993)
concerning low-mass-star-forming clumps is that they are
quasi-static equilibrium configurations with so-called
``subcritical'' mass-to-magnetic-flux ratios, so that the clumps
are supported against their self-gravity by the magnetic field in
the direction perpendicular to it, and by a combination of thermal
and turbulent pressures along the field. Under ideal MHD
conditions, the magnetic field is ``frozen'' into the plasma, and the
magnetic flux is conserved. Under the additional assumption that the
clump's mass is also constant, then the mass-to-flux ratio
is a fixed parameter of the clump, which therefore cannot collapse if
this ratio is subcritical
(meaning that the core's
self-gravity is never enough to overwhelm the magnetic support).
However, because the cold molecular gas is only partially ionized,
the process known as ambipolar diffusion causes a loss of magnetic
flux from the clumps on time scales long compared to their
free-fall time, allowing them to contract and form denser cores
that will ultimately collapse.

However, it well known that molecular clouds are supersonically
turbulent (e.g., Larson 1981; Blitz \& Williams 1999), and it is
becoming increasingly accepted that the cores within them are the
density fluctuations induced by the turbulence (\BP, \VS\ \& Scalo
1999; see also the reviews by \VS\ et al.\ 2000; Mac Low \&
Klessen 2004). In this context, the cores have a highly dynamical
origin (supersonic compressions), and their masses are hardly
fixed. Moreover, it is natural to ask whether they can settle into
hydrostatic equilibria, an event which requires the equilibria to
be stable (or ``attracting'', in the language of nonlinear
phenomena). Otherwise, the dynamic density fluctuations will just
``fly past'' the equilibrium state on their way to collapse, or
else ``rebound'' and merge back with their environment, if they do
not quite reach the equilibrium point. In this case, the cores'
lifetimes should be much shorter than in the standard model,
probably comparable to their free-fall times.
In the present contribution, we argue in favor of this scenario. To
this end, we discuss the formation and evolutionary time scales of
cores that form in numerical simulations of isothermal,
compressible MHD turbulence, in relation to the magnetic
criticality of the whole computational box. Here we present a
brief overview. For a full, detailed discussion, see \VS\ et al.\
(2004).

\section{Qualitative considerations} \label{sec:quali}

For the discussion below, it is convenient to characterize a core
by two non-dimensional parameters: its {\it Jeans number}
$\Jc=R_{\rm c}/L_{\rm J}$, giving the ratio of the core's radius
$R_{\rm c}$ to its Jeans length $L_{\rm J}$, so that a
gravitationally unstable core (with respect to thermal support)
has $\Jc > 1$, and its mass-to-flux ratio $\muc$, normalized to
the critical value for magnetic support, so that a supercritical
structure has $\muc > 1$. The corresponding values for the entire
computational box are denoted $J$ and $\mu$.

To assess the necessary conditions for the formation of either
sub- or supercritical cores, we note that the mass-to-flux ratio
of an {\it isolated} cloud constitutes an upper bound for that of any
subregion within it, since, as long as the flux-freezing
condition holds, both the cloud's mass and magnetic flux are
fixed, regardless of the cloud's volume. So, even if the entire
mass of the cloud were compresed into the core's volume, its
mass-to-flux ratio would remain the same. Thus, in the absence of
ambipolar diffusion, supercritical cores can only form within
supercritical clouds or clumps. On the other hand, subcritical
cores can arise in either sub- or supercritical clouds. Now, a
{\it gravitationally bound}, yet subcritical clump must have $\Jc
> 1$ in addition to $\muc <1$, so that it does not re-expand after
the turbulent compression that formed it subsides. However, recent
observational studies of the magnetic field strengths in molecular
clouds suggest that these are the objects with highest
mass-to-flux ratio in the hierarchy going from diffuse clouds to
dense cores (e.g., Crutcher 2004), and are likely to be
supercritical in general (e.g., Bourke et al.\ 2001), so the
formation of supercritical cores without the need for ambipolar
diffusion appears feasible in real molecular clouds.

\section{Numerical simulations and Results} \label{sec:num}

In order to investigate the magnetic criticality of the cores that
can form in turbulent environments in relation to that of their
parent clouds, we have performed a suite of numerical simulations
of turbulent flows at fixed rms sonic Mach number ($\Ms \sim 10$)
and {\it global} Jeans number $J=4$, and varying the plasma
$\beta$, defined as the ratio of thermal to magnetic pressure, in
order to consider cases that are subcritical (denoted $\beta$.01,
with $\mu = 0.9$), mildly supercritical (denoted $\beta$.1, with
$\mu = 2.8$), strongly supercritical (denoted $\beta$1, with $\mu
= 8.8$), and non-magnetic (denoted $\beta$$\infty$, with $\mu =
\infty$). These simulations can be thought of as representing
regions of size $L=4$ pc, with a mean number density $n_0 = 500$
cm$^{-3}$, a turbulent velocity dispersion of 2 km s$^{-1}$, and a
sound speed $c_{\rm s} =0.2$ km s$^{-1}$, 
with mean field strengths of 46, 14.5, 4.5 and 0 $\mu$G,
respectively.

\begin{figure}[!t]
  \includegraphics[width=\columnwidth,height=5cm]{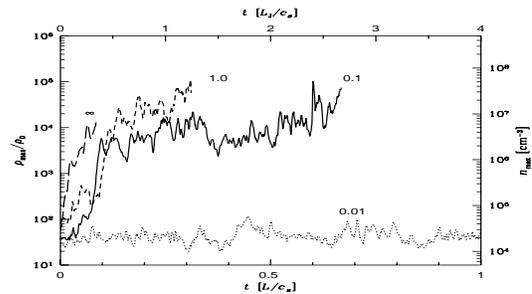}
  \caption{Evolution of the global maximum of the density field for
all runs 
considered here. The numbers indicate
the corresponding values of $\beta$.}
  \label{fig:rho_max}
\end{figure}

Figure \ref{fig:rho_max} shows the evolution of the global density
maximum for all four runs, with the time axis shown in units of
the sound crossing time $\ts \equiv L/c_{\rm s}= 20$ Myr (lower axis)
and of the global free-fall time $\tfg \equiv L_{\rm J}/c_{\rm s} = 5$
Myr (upper axis). It is 
readily seen that the subcritical run $\beta$.01 does not produce
very large density enhancements, with the global density maximum at
any given time being $\sim 30 n_0$, and never exceeding $100 n_0$. These
values are too low, and the transients are too short ($< 1$ Myr),
for ambipolar diffusion to operate and reduce the magnetic flux
under canonical estimates of the ambipolar diffusion time scale
(see, e.g., McKee et al.\ 1993). Note also that this occurs even
though this run is very close to global criticality, and thus the
production of gravitationally-bound cores in subcritical
environments appears to be highly unlikely even for nearly
critical boxes. Finally, note that the densities encountered in
this run are also safely below the ``Jeans criterion'' of Truelove
et al.\ (1997) ($n<256 n_0$ for our resolution of $256^3$ grid
zones) and the extension of it proposed by Heitsch, Mac Low \& Klessen (2001)
($n<115 n_0$) to ensure that numerical diffusion does not
significantly damp MHD waves within the cores, and so our result
is robust. Note, moreover, that these criteria may be excessive
for the problem at hand, as it was shown by Heitsch et al.\ (2001)
that the magnetic field can only prevent collapse if it provides
{\it magnetostatic} support, rather than wave-pressure support. So
our main resolution concern is to avoid magnetic flux loss, not
damping of MHD waves. In turn, the Truelove et al.\ (1997)
criterion is only defined in order to prevent artificial
fragmentation of collapsing structures, but here we are not
concerned with the internal structure of the cores as they
collapse; rather, we are just interested in whether collapse
occurs or not. Thus, a more appropriate criterion may be that the
internal mass-to-flux ratio of the cores does not exceed that of
the numerical box, or of its parent structure, as discussed above
(i.e., that the mass-to-flux ratio does not increase inwards of a
density structure).

In contrast, the supercritical and nonmagnet\-ic runs rapidly develop
densities approaching $10^4 n_0$, which correspond to collapsed,
unresolved objects. These densities occur at $\sim 1/2 \tfg$ in the
magnetic runs, and at $\sim 1/5 \tfg$ in the non-magnetic one. However,
inspection of an\-i\-mations of the simulations (available at {\tt
http://www. astrosmo.unam.mx/$\sim$e.vazquez/turbulence\_HP/
movies/VKSB04.html}, or in \VS\ et al.\ 2004)
shows that the typical time spans from the beginning of the turbulent
compression to the completion of collapse are $\sim$ 1--1.5 {\it local}
(i.e., at the mean density of the core) free-fall times ($\tfc$), or,
equivalently, $\sim$ 0.1--0.15 $\tfg$.

\begin{figure}[!t]
  \includegraphics[width=0.5\columnwidth,height=3cm]{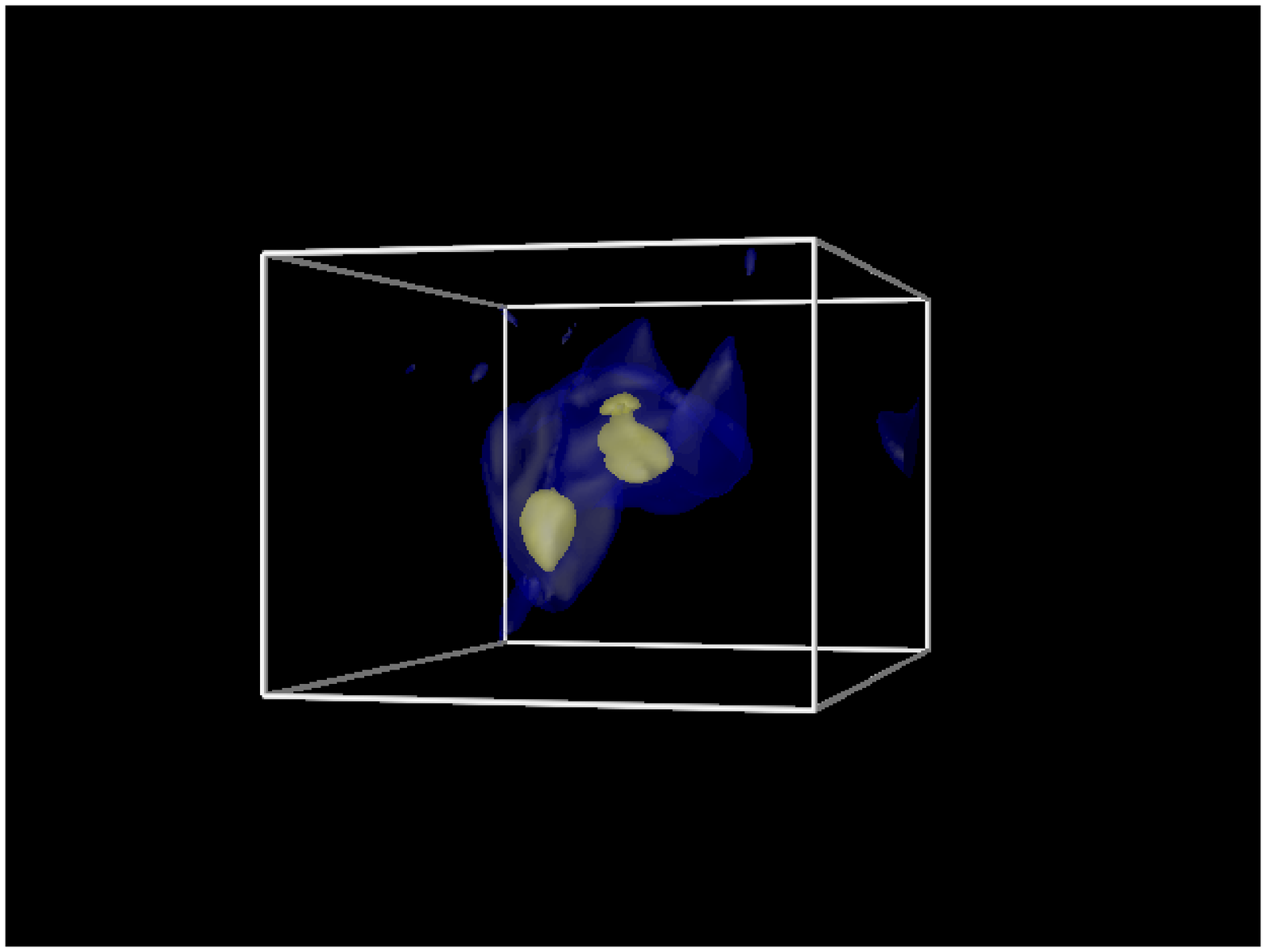}%
  \hspace*{.1\columnsep}%
  \includegraphics[width=0.5\columnwidth,height=3cm]{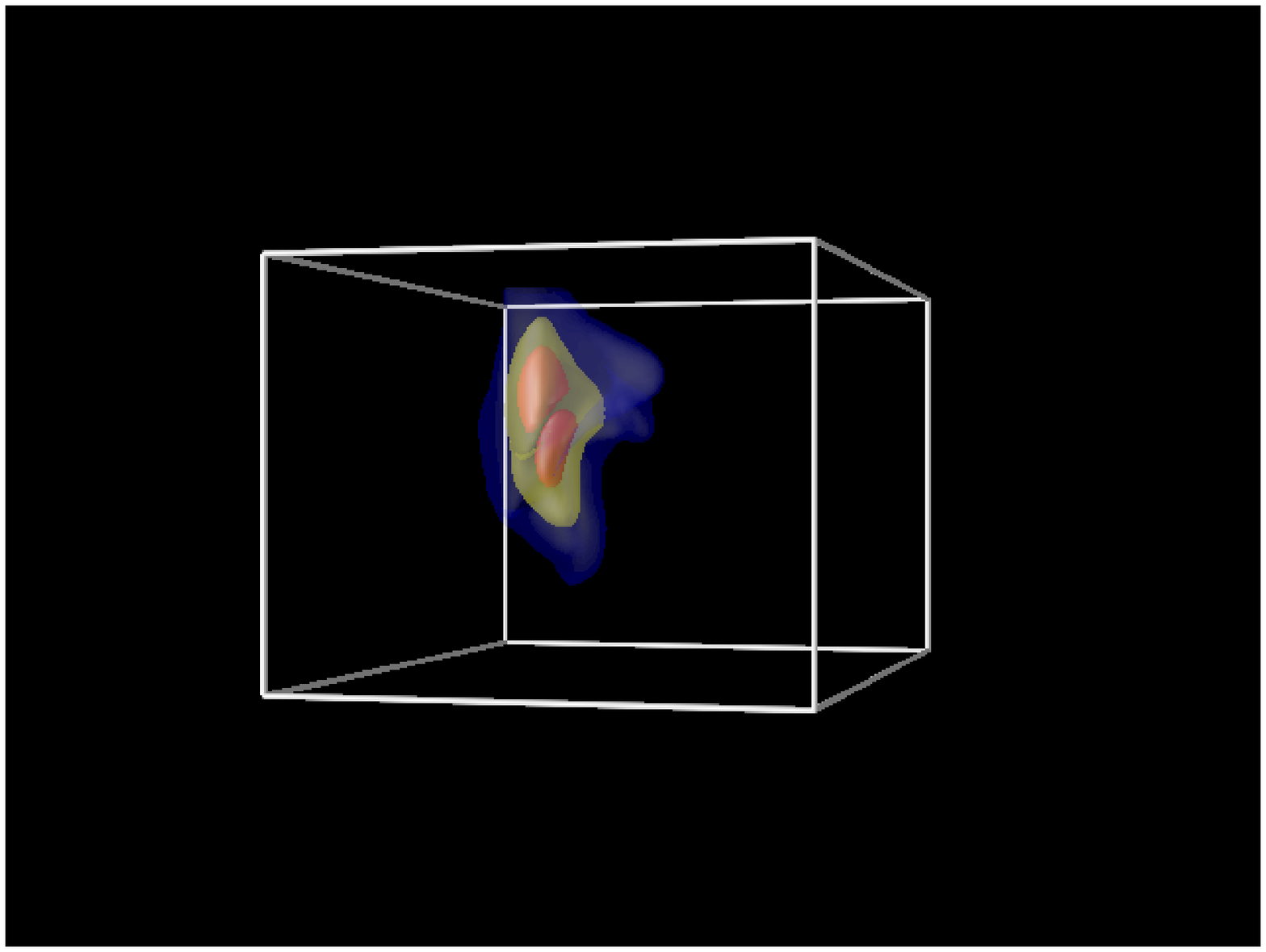}
  \caption{Iso-density surface 3D map of the clump-core (cores 1 and 2)
system appearing at early times in run $\beta$.1, detailed in
Table \ref{tab:cores_b.1}. The times shown are $t= 0.06 \ts$ ({\it
left}) and $t= 
0.08 \ts$ ({\it right}). The iso-density surfaces are at $n= 10
n_0$ (blue), $n= 40 n_0$ (yellow) and $n = 100 n_0$ (red). The
latter does not exist at $t= 0.06 \ts$.}
  \label{fig:cores_b.1}
\end{figure}

Measurement of the Jeans number and mass-to-flux ratio of the
cores as they are formed and evolve towards collapse sheds light
on how the process occurs. In Table \ref{tab:cores_b.1}, we present these
parameters, together with other relevant data, for the first
collapsed object that forms in run $\beta$.1. Other cores have
similar histories. The animation shows that this object forms out
of a larger clump initially containing two cores, which ultimately
merge to form the collapsed object (see also Figure
\ref{fig:cores_b.1}). Table \ref{tab:cores_b.1} thus gives the data for the
structures defined out to a threshold density level $\nt$ from the
local maxima at three different times during the evolution of the
system. The times are in units of the box sound-crossing time. At
the time $t=0.06 \ts$, the threshold $\nt=40 n_0$ resolves the two
cores, but at the two later ones only one core is seen above this
threshold (see Figure \ref{fig:cores_b.1}).

\begin{table*}[!t]\centering 
  \setlength{\tabnotewidth}{0.9\textwidth}
  \setlength{\tabcolsep}{1.33\tabcolsep}
  \tablecols{8}
\caption{Parameters of clumps and cores in run
$\beta$.1.} 
\label{tab:cores_b.1}
  \begin{tabular}{cccccccc}
    \toprule
Core name & \multicolumn{1}{c}{Time}\tabnotemark{a} &
\multicolumn{1}{c} {$\nt$}\tabnotemark{b}       &
\multicolumn{1}{c}{$\navg$}\tabnotemark{c}      &
\multicolumn{1}{c}{$\npeak$}\tabnotemark{d} &
\multicolumn{1}{c}{$R_{\rm c}/L$}\tabnotemark{e}&
\multicolumn{1}{c}{$J_{\rm c}$}\tabnotemark{f}  &
\multicolumn{1}{c}{$\mu_{\rm c}$}\tabnotemark{g}    \\
    \midrule

Parent clump    & 0.06  & 10    & 19.0  & 83.0  & 0.071 & 1.27  & 1.73  \\
Core 1      & 0.06  & 40    & 51.1  & 83.0  & 0.018 & 0.53  & 0.90  \\
Core 2      & 0.06  & 40    & 44.4  & 56.0  & 0.020 & 0.54  & 0.86  \\
Single core     & 0.08  & 40    & 78.5  & 337.  & 0.041 & 1.45  & 2.09  \\
Single core     & 0.10  & 40    & 329.  &$5.07\times 10^3$& 0.031& 2.28& 4.66\\

    \bottomrule

\tabnotetext{a} {Time of observation of structure, in units of $\ts$.}
\tabnotetext{b} {Threshold density for clump/core identification.}
\tabnotetext{c} {Mean density of clump/core.}
\tabnotetext{d} {Maximum density within clump/core.}
\tabnotetext{e} {Clump/core radius in units of the computational box
size $L$.}
\tabnotetext{f} {Ratio of core's radius to its Jeans length.}
\tabnotetext{g} {Mean mass-to-flux ratio within clump/core in units of
the critical value (compare to simulation value of 2.8).}

  \end{tabular}
\end{table*}

\ From Table \ref{tab:cores_b.1} it is seen that the parent clump
(defined by setting 
$\nt = 10 n_0$) is already super-Jeans and supercritical at
$t=0.06 \ts$, but the daughter cores (defined by $\nt = 40 n_0$)
are still sub-Jeans and subcritical. Both the mean and peak
densities (respectively $\navg$ and $\npeak$) within the parent
and daughter structures are safely below even the most stringent
condition of $n < 115 n_0$ for avoidance of MHD wave damping by
numerical diffusion, and so the supercritical nature of the parent
clump is a robust physical result, and the clump cannot be
supported by the magnetic field. Of course, there remains the
possibility of turbulent support, but since the clump itself was
formed by a turbulent compression, the compressive energy clearly
overwhelms the internal, random, supporting one, and in fact is
the mechanism that drives the clump into becoming super-Jeans and
supercritical. In any case, a detailed analysis of the virial
balance of the clump/core system and the role of the velocity
field, to be presented elsewhere, is necessary, but here we have
shown that the magnetic field is insufficient for supporting the
parent clump, even though its densest regions (the daughter cores)
are initially subcritical and sub-Jeans. At the later times $t =
0.08 \ts$ and $t = 0.1 \ts$, the single core defined by $\nt=40$
also has become super-Jeans and supercritical, as expected for the
generalized collapse of the parent clump. Although finally at time
$t=0.1 \ts$  numerical diffusion is clearly important ($\muc > \mu$),
this is seen to occur {\it after} the onset of the clump's
collapse, which began while the core/clump system was still well
resolved.

A final observation that here we only mention briefly (see \VS\ et
al.\ 2004 for a full discussion) is that in the strongly
supercritical run $\beta$1, two cores with slightly longer durations
($\sim 5 \tfc$) are formed, but in both cases they end up
re-dispersing, rather than collapsing. This is in agreement with the
simple argument that re-expansion should take longer times than
compression or collapse, because in that case self-gravity acts as
{\it retarding} agent against the re-expansion.

\section{Conclusions} \label{sec:conclusions}

In this contribution, we have presented a discussion and numerical
simulations of the formation, nature, and lifetimes of dense cores
in magnetized clouds. We argued that the mass-to-flux ratio of
a cloud puts an upper bound to that of any clump or core within it,
and so the 
only way to form supercritical cores is within supercritical
clouds (neglecting ambipolar diffusion). Moreover, numerical
simulations of marginally subcritical clouds show that no
gravitationally bound cores form in this case, while in
supercritical clouds the gravitationally bound cores that form
occur inside clumps that are supercritical, and rapidly become
supercritical themselves and collapse, in timescales that do not
exceed twice their local free-fall time $\tfc$. We also showed
that the mass-to-flux ratio of structures defined through a
density threshold (as would be the case of cores observed in a
single molecular line, which requires the density to be larger
than a certain value in order to be excited) does not remain
constant, because the core is continuously connected to its parent
structure, from which it can accrete mass.
Our results support the notion that clumps and cores are
out-of-equilibrium, transient structures, and that a class of
``failed'' cores should exist, that will not form stars.

\acknowledgements


\end{document}